# Synthesis of functional nitride membranes using sacrificial water-soluble BaO layers

Shengru Chen,[1,2] Qiao Jin,[1] Shan Lin,[1] Haitao Hong,[1,2] Ting Cui,[1,2] Dongke Rong,[1] Guozhu Song,[3] Shanmin Wang,[3] Kuijuan Jin,[1,2,5,*] Qiang Zheng,[4,*] and Er-Jia Guo[1,2,5,*]

**Affiliations**

[1] Beijing National Laboratory for Condensed Matter Physics and Institute of Physics, Chinese Academy of Sciences, Beijing 100190, China

[2] School of Physical Sciences and Center of Materials Science and Optoelectronics Engineering, University of Chinese Academy of Sciences, Beijing 100049, China

[3] Department of Physics, Southern University of Science and Technology, Shenzhen 518055, China

[4] CAS Key Laboratory of Standardization and Measurement for Nanotechnology, CAS Center for Excellence in Nanoscience, National Centre for Nanoscience and Technology, Beijing 100190, China

[5] Songshan Lake Materials Laboratory, Dongguan, Guangdong 523808, China

*Corresponding authors. Emails: kjjin@iphy.ac.cn, zhengq@nanoctr.cn, and ejguo@iphy.ac.cn

**Abstract**

Transition metal nitrides (TMNs) exhibit fascinating physical properties that show great potential in future device applications. Stacking two-dimensional TMNs with other functional materials with different orientations and symmetries requires separating epitaxial TMNs from the growth substrates. However, the lattice constants of TMNs are incompatible with those of most sacrificial layers, resulting to a great challenge in fabricating high-quality single-crystalline TMN membranes. In this study, we report the application of a water-soluble BaO sacrificial layer as a general method for creating freestanding TMN membranes. Using CrN as an example, the relatively small lattice mismatch and identical cubic structure between BaO and CrN ensure the formation of heterostructures. We directly observe the planar atomic structure and correlate its electronic state with its intrinsic transport properties using millimeter-size CrN membrane. Our research enables the fabrication of freestanding TMN membranes and transfer them to arbitrary substrates. By integrating TMN membranes with other materials will stimulate further studies in the emergent phenomena at heterointerfaces.

# Ⅰ. Introduction

Exfoliating single-crystalline thin films from growing substrates has recently undoubtedly had a significant impact in applied physics, including flexible displays, electronic "*skin*," etc. [1]. Recently, many efforts have focus on freestanding oxide membranes. Acid-soluble $La_{1-x}Sr_xMnO_3$, $SrCoO_{2.5}$, and $AlO_x$ [2-4] sacrificial layers have been used to synthesize oxide membranes. However, the strong chemical acid can afffect the sample quality, completeness, and physical properties of functional oxide membranes. Alternatively, Lu *et al.* reported a water-soluble $Sr_3Al_2O_6$ (SAO) as a sacrificial layer to eliminate acid and maintain high-quality membranes [5]. This method enables the application of extremely large stretching stresses and bending strain gradients to these flexible oxide membranes. Emergent phenomena, such as strain-induced ferromagnetic metal-to-insulator transition in manganites, ferroelectricity in center-symmetric $SrTiO_3$, and the giant tetragonality and polarization in single unit freestanding $BiFeO_3$, were observed [6-8]. Another advantage of freestanding membranes is the feasibility of integrating these functional materials on silicon or other CMOS-compatible substrates [9, 10]. This method may enable non-volatile memory or logic computing applications in the future.

The investigation of flexible TMN membranes is rare, despite great success in the fabrication of oxide membranes. Transition metal nitrides (TMNs) are another class of important materials that have the potential to significantly enhance several device designs and functionalities. Analogous to transition metal oxides, TMNs exhibit various physical properties, such as piezoelectricity [11], ferromagnetism [12], ferroelectricity [13], thermoelectricity [14, 15], and superconductivity [16-18], which make them appealing in device structures. The intriguing properties of TMNs and heterostructures have been systematically investigated for several decades. Over time, great success has been demonstrated in synthesizing stoichiometric single-crystal TMN thin films with low defect densities using magnetron sputtering and physical vapor deposition [12, 14, 19-23].

Among TMNs, CrN attracts special interest due to its high thermal stability, corrosion resistance, and thermoelectric properties [14, 15, 24]. It has a rock-salt structure with a lattice constant of 4.15 Å, which is compatible with most single-crystalline oxide substrates. CrN is a typical antiferromagnetic metal with a Néel temperature ($T_N$) of ~ 283 K [25-29]. The exceptional mechanical [30, 31] and transport properties [32-35] of CrN were heavily debated over decades. Taken electrical transport behaviors for instance, it is known that CrN is a paramagnetic metal at room temperature and transits into an antiferromagnetic metallic phase below $T_N$. The structural change from cubic (room temperature) to orthorhombic crystal structure is driven by the magnetic stress (below $T_N$) [28]. In the case of transport behavior, the phase transition accompanies a clear resistivity sudden jump at $T_N$. Extensive experimental works and theoretical calculations argue whether CrN is an intrinsic metal or insulator at room temperature. There are several controversial results on their intrinsic electronic states. Some groups showed that CrN presents a semiconductor-to-metal or insulator-to-insulator transition, whereas others confirmed that it stays metallic at all temperatures [36-42]. Several proposals have been made on the causes of the discrepant transport results for CrN thin films, such as ordered nitrogen vacancies, low crystallinity with impurity phases, surface conductivity, and substrate epitaxial constraints [37, 38, 43-48]. Recently, the theoretical analysis predicts that the ferromagnetic band structure of CrN is half metallic, indicating a strongly correlated behavior [49]. Furthermore, Yang et al. proposed that strong Dzyaloshinskii–Moriya interaction (DMI) could be induced in CrN monolayers with vertical electric polarization via the Rashba effect [50]. Potential applications using multiferroic skyrmions with electric-field switchable chirality and polarity based on CrN would provide a unique opportunity for future low-power devices.

To fabricate high-quality CrN films at high temperatures with *in-situ* compensation of nitrogen vacancies, our group used pulsed laser deposition with nitrogen plasma [51-53]. Using these methods, we can potentially combine compounds with different functionalities requiring different growth environments, such as oxides and nitrides [54]. We investigated the thickness and strain dependencies of structural and electrical transport properties of CrN thin films. Furthermore, we showed how to fabricate CrN freestanding membranes for the first time using SAO as sacrificial layers [51]. A direct comparison of structural and electronic states in strained thin films and freestanding CrN membranes was obtained. Unfortunately, we found that there are still some limitations with SAO sacrificial layers which make them unsuitable for CrN thin films. First, the fabrication of high-quality SAO is critical. SAO has a complex structure with 264 atoms in a huge unit cell ($a_{SAO}$ = 15.844 Å) [5]. Therefore, the growth conditions have a great influence on the quality of SAO, which consequently affects the succeeding layers. Secondly, the growth conditions for SAO and CrN are not compatible. The growth of SAO requires high oxygen partial pressure, whereas CrN needs to be grown in a vacuum. Therefore, the physical properties of TMN thin films will greatly deteriorate when they are accidentally oxidized. Thirdly, CrN films possess a huge lattice mismatch when grown on SAO. The lattice constant of SAO is smaller than four-unit-cells of CrN. The CrN films grown on SAO suffer ~–5% compressive strain, affecting the crystallinity of CrN films. To minimize the defect formation within CrN thin films, it is necessary to choose other sacrificial layers that can be grown in vacuum and have appropriate lattice constants.

Here, we show how to fabricate millimeter-scale freestanding CrN membranes using a water-soluble BaO layer. BaO layers can be grown on sapphire substrates with a primary orientation and dissolve in water rapidly. One of the advantages of using BaO instead of SAO as sacrificial layers is its wide-range tunable lattice constant that is compatible with most of TMNs. By substituting Ba with Sr, the lattice constant of (Ba, Sr)O layer changes from 5.539 to 5.161 Å, [55, 56] tuning the chemical ratio between Ba and Sr. In this way, various TMN membranes can be fabricated in a high-quality manner with wisely choice of lattice-matching sacrificial layers. We show that the millimeter-size CrN membranes are released from sapphire substrates after immersion in water. By transferring a small portion of membranes onto the TEM grid, the microstructure of CrN membranes is directly observed. After the removal of the substrate, the intrinsic electrical properties of CrN membranes are measured, yielding a bulk-like phase transition at decreasing temperatures.

**II. Experiments**

CrN/BaO bilayers were grown on (0001)-oriented sapphire substrates using pulsed laser deposition. Figure 1(a) shows a schematic of thin film deposition. The ceramic targets were ablated using a pulsed XeCl excimer laser with a wavelength of 308 nm, a repetition of 3 Hz, and a duration of ~ 25 ns. A $BaO_2$ pallet with a diameter of 1 inch was used as a target for BaO thin film growth; whereas a CrN target prepared using a high-pressure reaction route was used for CrN thin film growth. At first, an approximately 30 nm thick BaO layer was fabricated at a substrate temperature of 550°C under a base pressure of ~ 1 × 10$^{-8}$ Torr. The as-grown films were *in-situ* annealed for 1 h to form the correct stoichiometry. [57] Subsequently, a CrN layer with a thickness of approximately 20 nm was grown on BaO under the same conditions. The growth rates for BaO and CrN layers are approximately 60 and 85 pulses/unit cell, respectively. The film thicknesses were estimated by counting laser pulses. Previously, it is known that the lattice mismatch between the sacrificial layer and target materials is an important factor in fabricating high-quality freestanding membranes. Figure 1(b) shows the growing relationship between BaO and CrN layers. Both materials have a cubic lattice structure. The lattice constants of BaO and CrN are 5.539 and 4.15 Å, respectively. Direct cubic-on-cubic epitaxy will result in an extremely large lattice mismatch. However, when the unit cell of CrN is rotated in a plane along (001) orientation by 45°, the lattice mismatch between them will dramatically reduce, providing a strong basis for the epitaxial growth of CrN on BaO. As shown in Figure 1(c), BaO and CrN layers exhibit (00$l$) diffraction peaks, suggesting that all layers are well aligned along the out-of-plane direction.

### III. Results and discussions

Following the completion of CrN/BaO bilayers epitaxial growth and basic structural characterization, polydimethylsiloxane (PDMS) was mechanically pressed firmly on the as-grown samples. The PDMS coating is a supporting layer to prevent fragile CrN membranes from cracking or curling after releasing from rigid substrates. The sample with PDMS support is immersed in water for a few minutes. The sacrificial BaO layer gradually dissolves, and CrN freestanding membrane is released from substrates. Figure 1(d) shows the XRD θ-2θ scan of a freestanding CrN membrane. The inset of Figure 1(d) shows reciprocal space mapping of CrN membranes around the (002) diffraction peak. After the removal of substrates, the CrN layers release the compressive strain. The lattice constant of the CrN membrane decreases slightly to 4.146 Å, close to its bulk value (4.15 Å), indicating that the epitaxial strain from both substrates is completely removed. Furthermore, no XRD signals were detected from BaO, suggesting that no BaO remained in the freestanding CrN membranes.

Previously, the microstructural characterizations of as-grown TMN films or heterostructures are commonly examined by cross-sectional imaging using scanning transmission electron microscopy (STEM). The complex preparation process may inevitably damage specimens during ion milling and mechanical thinning. Thicknesses of specimens also affect the quality of STEM imaging. Therefore, we directly transferred a 20-nm-thick freestanding CrN membrane onto a copper grid for STEM observations to avoid these damages during specimen preparation, as the release and transferring processes shown in Figures 2(a)-2(c). STEM observations were performed on a Double-Cs-corrected transmission electron microscope (Spectra 300, Thermo Fisher Scientific), operated at an accelerating voltage of 300 kV. High-angle annular dark-field (HAADF)-STEM images were collected with a probe convergence angle of 25 mrad and collection angles that ranges from 62 to 200 mrad. The inset of Figure 2(d) shows a low-magnification HAADF-STEM image for a tiny portion of CrN membranes on a specific mesh of copper grid. The thickness of CrN membranes is ~ 20 nm, providing a great opportunity to directly image the crystal structure from the sample's top without involving a complex specimen preparation process. Figure 2(d) displays a mid-magnification plane-view HAADF-STEM image of CrN membranes. The white spots represent the positions of Cr atoms, N atoms was not observed due to the overlap with Cr atoms. We note that the crystalline structure of CrN is well preserved and no clear structural defects in membrane form are observed after release and transfer processes. An enlarged atomic-scale HAADF-STEM image in Figure 2e shows a representative structure of the CrN lattice, where Cr atoms are precisely marked in blue spheres. The average distance between two Cr atoms is measured to be 2.08 ± 0.06 Å, corresponding to the half-length of a CrN single unit cell.

The electronic state of freestanding CrN membranes was further investigated by electron energy-loss spectroscopy (EELS), which highlights the atomic configuration and bonding states of both Cr and N. EELS data were acquired using a Gatan image filter (Gatan Inc.) with a dispersion of 0.6 eV per channel and collection semi-angle of 40 mrad. Figure 3(a) shows an elemental-specific electron energy-loss (EEL) spectrum covering both N $K$- and Cr $L$-edges. We compare the peak positions of Cr $L$-edges to the reference EEL or XPS spectra for $Cr^{3+}$ ions obtained from $Cr_2O_3$ [58-60] and stoichiometric CrN thin films [61], suggesting the valence state of Cr ions in freestanding CrN membranes is +3. STEM and EELS results indicate that freestanding CrN membranes maintain the crystal structure, negligible defects, and nondetectable chemical disorders.

The electrical property of CrN is another indicator of structural disorder and chemical uniformity. Figure 3(b) shows the resistivity as a function of temperature for a freestanding CrN membrane attached to a PDMS support. The inset shows the schematic of the measuring setup using a two-point-contact configuration. We observe a clear resistivity drop at ~ 200 K. We believe this insulating phase transits to the metallic phase, corresponding to the structural and magnetic phase transition of CrN. Compared to the earlier reports on epitaxial CrN films with identical film thickness grown on MgO and $Al_2O_3$ substrates [51-53], transition temperature enhances over 100 K. Apparently, the increase of transition temperate is subject to the removal of epitaxial constraints. Nevertheless, it is worth stressing that the observed transition temperature is still ~ 80 K lower than that of bulk CrN and our earlier reports. We believe at least three major factors influence the reduced transition temperature. First, strain effects due to the natural folding after removal from the substrate. Previously, we had shown that the critical thickness for metal-to-insulator transition in CrN is ~ 30 unit cells. In this study, the thickness of CrN membranes is 20 nm at the boundary of the bandgap opening. Small perturbations may lead to significant changes in electronic properties. Secondly, the inevitable defects present in the membranes during the releasing process. These defects will heavily scatter the conducting carriers resulting in enhanced resistivity. Thirdly, the nitrogen vacancies during thin film deposition. In this work, CrN films were grown under reduced conditions, e. g., in a vacuum. Unlike our previous work, active nitrogen plasma was introduced to compensate for the nitrogen-vacancy in CrN films. Therefore, CrN membranes studied in the present work may contain more nitrogen vacancies than CrN films in our previous work. This fact is directly proven by observing two orders of magnitude enhanced resistivity in our freestanding CrN membranes compared to earlier work. In addition, we also want to mention that the surface scattering of charge carriers affects the transport behavior of thin films. The flow of carriers would be constrained in the thin films/membranes as their thickness reduces. It, in turn, may contribute to the observed resistivity enhancement in CrN freestanding membranes.

## IV. Conclusions

We demonstrate a feasible means for fabricating TMN membranes using a water-soluble sacrificial BaO layer. Using CrN as an example, we show that a freestanding CrN membrane maintains its lattice structure without apparent defects/dislocations after releasing and transferring processes. The CrN membranes exhibit a distinct insulator-to-metal transition at Néel temperature, which is substantially higher than that of a strained epitaxial film, without the substrate's constraint. Our work shows a straightforward method for fabricating freestanding membranes that can only be grown in a reduced condition, for instance, in a vacuum. Furthermore, using a sacrificial BaO layer also provides a choice for freestanding functional materials that naturally have large lattice constants. The sacrificial BaO would greatly reduce the lattice mismatch and benefit a highly epitaxial thin film growth.

**Acknowledgements**

This work was supported by the National Key Basic Research Program of China (Grant Nos. 2020YFA0309100 and 2021YFA1202801), the National Natural Science Foundation of China (Grant Nos. 11974390 and 11721404), the Beijing Nova Program of Science and Technology (Grant No. Z191100001119112), the Guangdong-Hong Kong-Macao Joint Laboratory for Neutron Scattering Science and Technology, and the Strategic Priority Research Program (B) of the Chinese Academy of Sciences (Grant Nos. XDB33030200 and XDB36000000).

**Figures and figure captions**

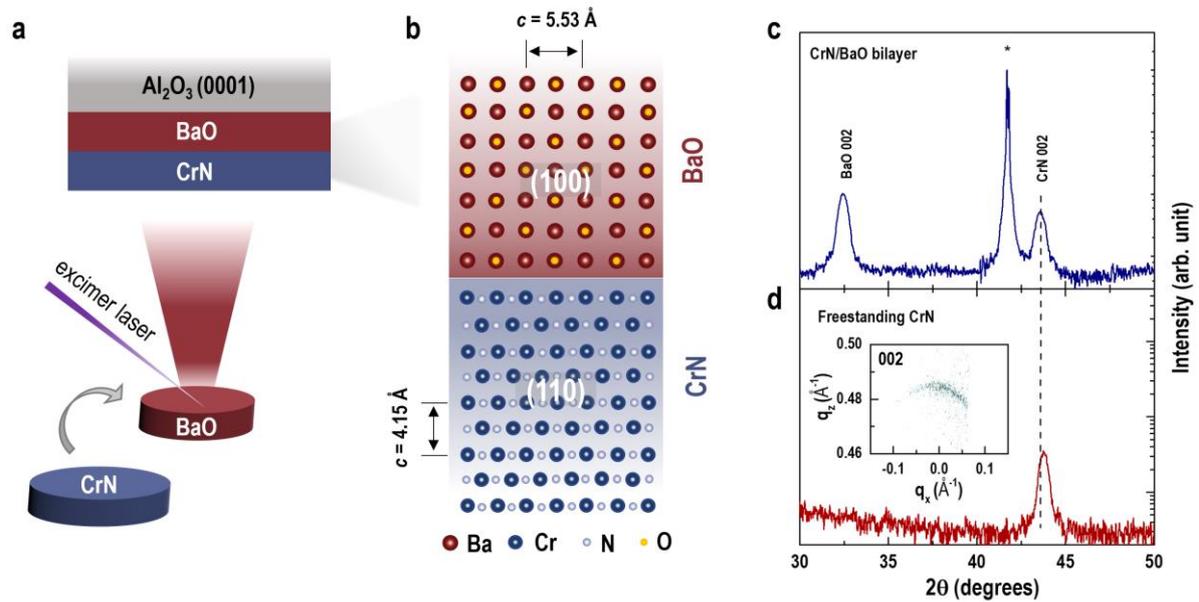

**Figure 1. Fabrication of freestanding CrN membranes.** (a) Growth of CrN/BaO bilayer consequently using pulsed laser deposition technique. (b) Schematic diagram of crystal structure across the CrN/BaO interfaces. Both layers are (00*l*) oriented. The unit cell of CrN is rotated in a plane along (001) orientation by 45° with respect to that of BaO, resulting in the (110) plane of CrN being parallel to the (100) plane of BaO. (c) and (d) XRD θ-2θ scans of a CrN/BaO bilayer on sapphire substrates (indicated with *) and a freestanding CrN membrane, respectively. The inset of (d) shows the reciprocal space map of the CrN membrane around the 002 diffraction peak.

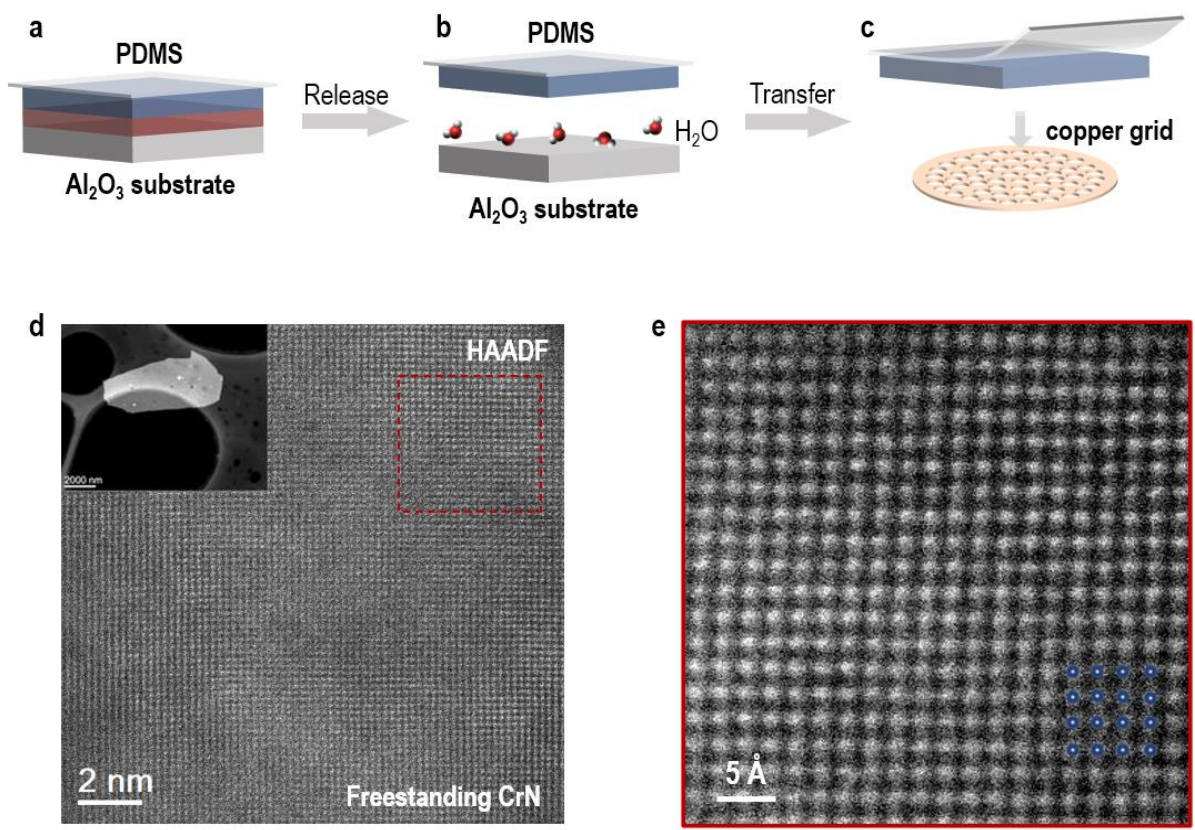

**Figure 2. Planar-view STEM imaging of freestanding CrN membranes.** (a)-(c) Fabrication process of CrN membrane specimens for STEM imaging. (d) Planar-view HAADF-STEM image of a freestanding CrN membrane. The inset shows a low-magnification HAADF-STEM image of a freestanding CrN membrane on a copper grid. (e) A zoom-in HAADF-STEM image from a representative region marked in a red dashed square in (d). Atomic positions of Cr ions are marked in (e).

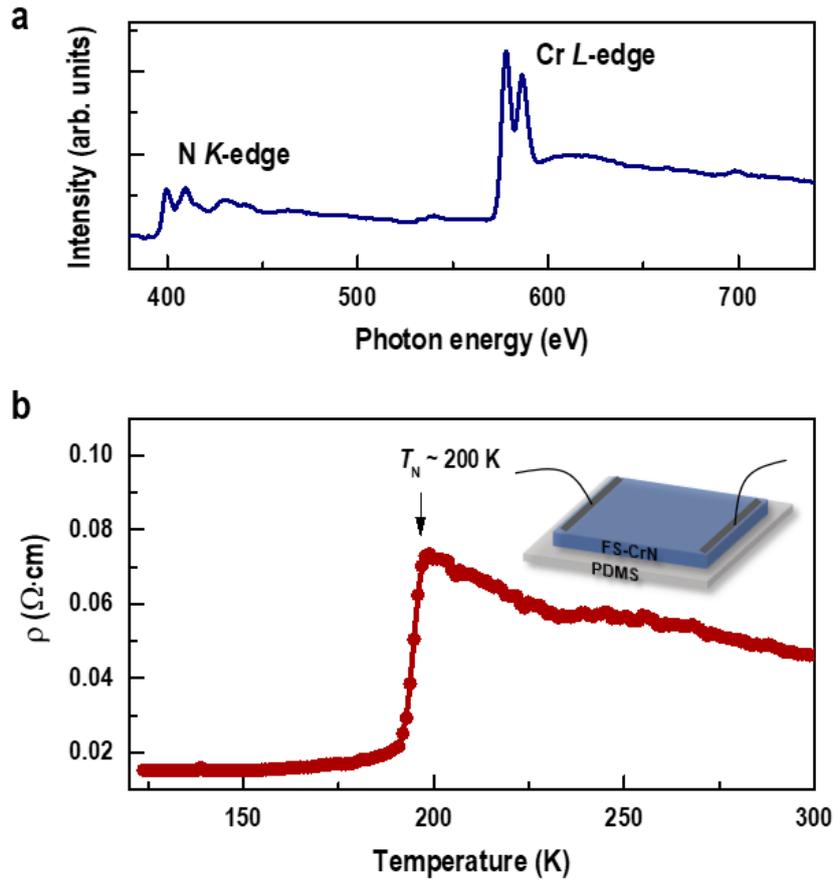

**Figure 3. Electronic state and transport properties of a freestanding CrN membrane.** (a) Electron energy-loss (EEL) spectrum for N *K*- and Cr *L*-edges obtained from a freestanding CrN membrane. The spectrum was averaged over an inspected region, revealing the presence of both Cr and N in the membranes. (b) Temperature dependent resistivity of a freestanding CrN membrane. The sharp resistivity drop at ~ 200 K indicates the structural and magnetic phase transition as decreasing temperature.